\documentclass[12pt]{article}

\begin{document}
\setlength{\textheight}{23cm}
\setlength{\topmargin}{-1cm}
\renewcommand{\thefootnote}{\fnsymbol{footnote}}
\font\vmath=msbm10 at 12pt
\def\vmi#1{\hbox{\vmath #1}}
\def\bmi#1{\mbox{\boldmath $#1$}}
\def\sp{\phantom{a}}

\begin{titlepage}
\font\csc=cmcsc10 scaled\magstep1
{\baselineskip=14pt
 \rightline{
 \vbox{\hbox{hep-th/0307236}
    }}}

 \renewcommand{\thefootnote}{\fnsymbol{footnote}}
 
     \begin{Large}
       \vspace{2cm}
       \begin{center}
         {Note on a fermionic solution of the matrix model and noncommutative superspace}      \\
       \end{center}
    \end{Large}

  \vspace{1cm}

\begin{center}
           SHIBUSA, Yuuichirou  \footnote
           {
e-mail address : shibusa@riken.go.jp} and
          TADA, Tsukasa  \footnote
           {
e-mail address : tada@riken.go.jp}
          
       \it Theoretical Physics Laboratory\\
          The Institute of Physical and Chemical Research (RIKEN)\\
          Wako 2-1, Saitama 351-0198, Japan

       \end{center}

\vfill

\begin{abstract}
We present a new  fermionic solution of the supersymmetric matrix model. The solution satisfies the commutation and anticommutation relations for noncommutative superspace. Therefore the solution can be considered as an implementation of noncommutative superspace on the matrix model.
\end{abstract}

\vfill
\end{titlepage}
\vfil\eject

\baselineskip=16pt
\setcounter{footnote}{0}
\renewcommand{\thefootnote}{\arabic{footnote}}

\section{Introduction}
Noncommutative space has been a fascinating arena in the study of non-perturbative aspects of string theory \cite{Connes:1997cr, Seiberg:1999vs, Aoki:1999vr, Douglas:2001ba}. There is now  a surge of renewed interest in noncommutativity, this time in superspace \cite{Klemm:2001yu,Ooguri:2003qp,Ooguri:2003tt,deBoer:2003dn,Kawai:2003yf,Chepelev:2003ga,Seiberg:2003yz,Britto:2003aj,Berkovits:2003kj,Hatsuda:2003ry,Terashima:2003ri,Ferrara:2003xy, Park:2003ku,Grisaru:2003fd,Britto:2003kg,Romagnoni:2003xt}. In this note, we study  noncommutative superspace as a fermionic background of a certain matrix model. 

In the so-called reduced model \cite{Eguchi:1982nm},  the information of spacetime is subtly converted into  large N color degrees of freedom. This  yields a wider perspective to the implementation of space-time in quantum field theory and manybody systems \cite{Parisi:1982gp,Gross:at,Bhanot:1982sh,Das:1982ux,Gonzalez-Arroyo:1982hz}. The crucial point is to respect the appropriate symmetries that  would correspond to the symmetries of the resultant field theory. Following the same spirit, one would be naturally lead to IIB matrix model \cite{Ishibashi:1996xs}:
\begin{equation}
S  = 
Tr({1\over 4}[X_{a},X_{b}][X^{a},X^{b}]
-{1\over 2}\bar{\theta}\Gamma ^{a}[X_{a},\theta ]) ,
\label{action1}
\end{equation}
where maximal 32 supersymmetries are realized on the set of $N \times N$ matrices.
In the above action (\ref{action1}), the contractions of the upper and lower indices are performed over 10-dimensional flat Minkowski metric. However, thanks to the maximal supersymmetry, the dynamics of spacetime is contained in the various configurations of the matrices. In fact, this model exhibits non-trivial spacetime backgrounds such as D-branes as a solution to the equation of motion from  (\ref{action1}). 

While one can investigate the IIB matrix model by the expansion around the commutative (diagonal) background \cite{Aoki:1998vn}, it is also interesting to start with the following solution:
\begin{equation}
[X^{a},X^{b}]=-i C^{ab} { \vmi{ I}}_{N}, \theta^{\alpha}=0,  \label{pqsol}
\end{equation}
where $C^{ab}$ is an anti-symmetric constant. The result is noncommutative Yang-Mills theory \cite{Aoki:1999vr}. $X_{a}$'s which satisfy eq. (\ref{pqsol}) themselves serve as noncommutative coordinates. In the string picture, this corresponds to the appearance of noncomuutative geometry in a constant B field background \cite{Seiberg:1999vs}.

One may regard the foregoing analysis as the representation of  (local) Poincar{\' e} group and its certain noncommutative version in the large N matrix model. Since superspace is nothing but the coset space of the super Poincar{\' e} group divided by the Lorentz group, it would be also natural to try to represent (noncommutative) superspace on the matrix model in a similar manner. In this letter, we exhibit such an attempt.

Also,  searching a solution of the matrix model that corresponds to the graviphoton background considered in \cite{Ooguri:2003qp,Ooguri:2003tt,deBoer:2003dn,Seiberg:2003yz} would have its own prudence. Considering that the noncommutative solution (\ref{pqsol}) corresponds  to a B field background in string theory, noncommutative superspace realized in the matrix model should have some connection with the graviphoton background. 

In the following, we present a solution with non-zero fermion matrices for the matrix model.  $X_{a}$'s and $\theta^{\alpha}$'s in the solution satisfy the relations which are similar to  those recently studied in the context of noncommutative superspace. It will be shown that the solution preserves half the supersymmetries contained in the model and the killing spinor is explicitly constructed. 

To illustrate the basic idea for our fermionic solution, let us consider  following fermionic matrices;
\begin{equation}
\theta \sim (Grassmann) \otimes Q +  (Grassmann) \otimes P + \cdots,
\end{equation}
where $Q$ and $P$ are a pair of matrices such that $[Q,P] =i$ .
The anticommutation relations among them would be something like
\begin{equation}
\{ \theta, \theta \} \sim i (Grassmann)^{2} \otimes \vmi{I}.
\end{equation}
Then, as we will next describe in detail,  we find that there is a set of bosonic matrices of the following form;
\begin{equation}
X \sim (Grassmann)^{2} \otimes Q +  (Grassmann)^{2} \otimes P + \cdots
\end{equation}
which satisfy both (anti-)commutation relations and the equation of motion for the matrix model.


\section{ Noncommutative superspace and the four-dimensional matrix model}
For the sake of the recent interest, let us first consider the four dimensional case.
In four dimension, the following commutation and anticommutation relations for the supercoordinates has been studied  in \cite{Seiberg:2003yz} :
\begin{eqnarray}
\label{ferminoncom}
&&\{\theta^{\alpha},\theta^{\beta}\}=C^{\alpha\beta}\\
\label{boseferminoncom}
&&[X^a,\theta^{\alpha}]=iC^{\alpha\beta}\sigma^a_{\beta \dot{\alpha}}
\bar{\theta}^{\dot{\alpha}},\\
\label{bosenoncom}
&&[X^a,X^b]=(\bar{\theta})^2C^{ab}  \\
&&\{{\bar \theta}^{\dot \alpha},{\bar \theta}^{\dot \beta}\}=\{{\bar \theta}^{\dot \alpha},{ \theta}^{\beta}\} = [{\bar \theta}^{\dot \alpha}, X^{a}]=0, \label{fermibar}
\end{eqnarray}
where\begin{eqnarray}
a,b &=& 1 \dots 4, \sp \sp \alpha,\dot{\alpha}=1 ,2, \nonumber \\
\label{Cdef1}
C^{ab}&\equiv&C^{\alpha\beta}(-\sigma^{ab} \epsilon)_{\alpha\beta}, \\
\label{Cdef2}
C^{\alpha \beta}&=& (\epsilon\sigma^{ab})^{\alpha \beta}C_{ab}.
\end{eqnarray}
We have followed the notation of \cite{Wess:cp}.

To analyze these relations in the matrix model, we consider the following four-dimensional matrix model;
%
%
\begin{eqnarray}
\label{action}
S&=&Tr(\frac14 [X^a,X^b]^2+ {\cal C}\theta\sigma^a [\bar{\theta},X_a]),
\label{4Daction}
\end{eqnarray}
%
%
The constant $\cal C$ is left undetermined for the moment.  
We  treat the model in the Euclidean signature  while maintaining the notation of \cite{Wess:cp} just as has previously been done in \cite{Seiberg:2003yz}. This is achieved by defining $\sigma^{0}$ here as $i$ times that of  \cite{Wess:cp}.
Then  $\theta$ and $\bar \theta$ become 
 independent \cite{Osterwalder:1972} and the equations of motion are as follows:
\begin{eqnarray}
\label{motionX}
[X^b,[X^a,X_b]]+{\cal C}
\{ \theta ,\sigma^a\bar{\theta} \} &=& 0 \\
\label{motionTh}
[X_a, (\theta\sigma^a)_{\dot{\alpha}}]&=&0 \\
\label{motionbTh}
[(\sigma^a \bar{\theta})_{\alpha}, X_a]&=&0.
\end{eqnarray}
In particular, (\ref{motionTh}) can also be derived by multiplying $\sigma_{a}$ on the both side of (\ref{boseferminoncom}) since $\sigma^a_{\beta \dot{\alpha}}(\sigma_{a})_{\alpha \dot{\gamma}} = -2\epsilon_{\beta\alpha}\epsilon_{\dot{\alpha}\dot{\gamma}}$, which cancels with symmetric $C^{\alpha\beta}$. This suggests that the equation of motion of the matrix model (\ref{4Daction})
is compatible with the algebra of non-commutative superspace (\ref{ferminoncom})-(\ref{fermibar}).
 In the following, we will show that it is actually the case by constructing an explicit solution of the equations of motions that also satisfies  (\ref{ferminoncom})-(\ref{fermibar}).

It is worth mentioning in passing, though, that this matrix model has the following enhanced (${\cal N} = 2$) supersymmetry transformations:
\begin{eqnarray}
\label{susyX}
\delta X^a &=& -i \xi\sigma^a\bar{\theta}+i\theta\sigma^a\bar{\xi} \\
\label{susyTh}
\delta \theta^{\alpha}&=& {\cal A}[X_a,X_b](\xi\sigma^{ab})^{\alpha}+
\xi '^{\alpha} \vmi{ I}  \\
\label{susybTh}
\delta \bar{\theta}^{\dot{\alpha}}&=& {\cal B}[X_a,X_b](\bar{\sigma}^{ab}
\bar{\xi})^{\dot{\alpha}}+\bar{\xi}'^{\dot{\alpha}}\vmi{ I}.
\end{eqnarray}
Since $\theta$ and $\bar \theta$ are now being treated as independent, 
we have introduced here independent parameters ${\cal A} $ and $ {\cal B}$. 
To satisfy the consistency of the algebra  (\ref{susyX})-(\ref{susybTh}) and
ordinary SUSY algebra, it is necessary that  ${\cal A}=-{\cal B}$. This further leads the commutator of two supersymmetries as follows:
\begin{eqnarray}
[\delta_1,\delta_2]&=& \vmi{ I}(i\xi_1\sigma^a\bar{\xi}'_2 
                       +i\xi'_1\sigma^a\bar{\xi}_2)\partial_a + \delta_{U(N)gauge}(-2i{\cal A}X_c\xi_1\sigma^c\bar{\xi}_2) - (1\leftrightarrow 2) \nonumber\\
\label{susyalg}
&& + \mbox{eq. of motion for }  \theta \mbox{ and } {\bar \theta},
\end{eqnarray}
in a shorthand notation.
Using these relations, it is straightforward to show that the action (\ref{4Daction}) is invariant under (\ref{susyX})-(\ref{susybTh}) provided
\begin{eqnarray}
{\cal A}{\cal C}&=& -i. \label{AC}
\end{eqnarray}  
One can use the above relation (\ref{AC}) to determine $\cal C$ in terms of $\cal A$.

Now we proceed to find a solution of (\ref{motionX})-(\ref{motionbTh}) with non-zero fermion matrices. Let us denote $U(N)$ generators as $T^{\hat{A}}$ and choose a integer $n$ which is large enough, but much smaller than $N$ so that $N/n >>1$.  The reason for introducing $n$ will  shortly become clear.
We will focus our attention on the following special  generators which have $n$ by $n$ block structure:
\begin{eqnarray}
\hat{A}&=&0,A \nonumber \\
A&=&1,2, \dots , 2n \nonumber \\
T^A&=& Q_1,P_1,Q_2,P_2,\dots , Q_{n}, P_{n},
\end{eqnarray}
where $Q_{k}$'s and $P_{k}$'s are $N/n \times N/n$ matrices and  $Q_{k}$'s and $P_{k}$'s satisfy
\begin{equation}
[Q_{j}, P_{k}] = i \delta_{jk}.
\end{equation}
We also denote the identity as $T^0$, that is,
\begin{equation}
T^0= \vmi{ I} _{N}.
\end{equation}
It follows that the only nontrivial structure constants $f_{\hat{A}\hat{B}\hat{C}}$ among the generators we have introduced are
\begin{eqnarray}
f_{AB0}&=&
\left(
\begin{array}{cccccc}
 0  & i & 0 & 0  & 0 & 0\\
 -i  & 0 & 0 & 0 & 0 & 0\\
 0  &  0 & 0 & i & 0 & 0\\
 0  &  0 & -i & 0 & 0 & 0 \\
 0 & 0 & 0 & 0 & \cdot &   \\
 0 & 0 & 0 & 0 &  & \cdot
\end{array}
\right). 
\end{eqnarray}

We will try to find a new solution in terms of the above introduced generators, that is to say, we give an ansatz for the solution in the following form: 
\begin{eqnarray}
X_{sol} ^{a}&= &\sum_{A=1}^{2n} X^{aA} T^{A} \\ \label{xsol}
\theta_{sol} ^{\alpha}&= &\sum_{A=1}^{2n}\theta^{\alpha A} T^{A} ,  \quad {\bar \theta}_{sol} ^{\dot \alpha} = \ \sum_{A=1}^{2n}{\bar \theta}^{{\dot \alpha}A} T^{A} .\label{thetasol}
\end{eqnarray}
The equation of motions provides the following conditions for the ansatz:
\begin{eqnarray}
\label{solbth}
\bar{\theta}^{\dot{\alpha}\sp  0}&=& \bar{\theta}^{\dot{\alpha}}\\
\label{solX}
X^{aA}&=&-i\theta^{\alpha A}(\sigma^a)_{\alpha \dot{\beta}}\bar{\theta}^{ \dot{\beta}0}  \\
\mbox{Others}&=& 0. \label{solO}
\end{eqnarray}

These relations give a new solution with nonzero fermionic matrices.  Moreover, if we denote 
\begin{equation}
\label{solTh}
 \sum_{AB}\theta^{\alpha A}\theta^{\beta B}f_{AB0} \equiv C^{\alpha \beta},
\end{equation}
$X_{sol} $,  ${\theta}_{sol}$ and ${\bar \theta}_{sol}$ obey the noncommutative superspace commutative and anticommutative relations (\ref{ferminoncom}) - (\ref{fermibar}) provided that  (\ref{xsol}) - (\ref{solO}) hold.
In other words, we can reproduce the relations (\ref{ferminoncom})-(\ref{fermibar}) in the matrix model (\ref{4Daction}).

A few brief comments follows. In the above analysis, one could say that $C^{\alpha \beta}$ is rather defined by the lefthand side of (\ref{solTh}).  So we are not constructing a solution for a given $C^{\alpha \beta}$. There is a possibility of another solution for more general $C^{\alpha \beta}$ that is discussed later.
Also  $C^{\alpha \beta}$ is not genuine c-number but bi-Grassmann in our case. Since the square of Grassmann number equals zero, it might cause a problem. However,  note that $(C^{\alpha \beta})^{k} \neq 0$ provided $k \leq 2n$ and one can take $n$ as large as one wishes in the large $N$ limit.  This is because we have  $2n$ sets of Grassmann variables due to the  $n$ by $n$ block structure of the matrices. Therefore,  with the relation (\ref{solTh}), we can avoid the power of $C^{\alpha \beta}$ from vanishing to a certain extent.

Let us next discuss the symmetry of the solution. It turns out that this background Killing spinor can be defined by using a two-component spinor $\tilde{\xi}$ as follows:
\begin{eqnarray}
\label{kill1}
\xi &=& \tilde{\xi}(\bar{\theta}_{sol})^2 \\
\label{kill2}
\bar{\xi}&=&\bar{\tilde{\xi}}\epsilon_{\alpha_1\alpha_2}\epsilon_{\beta_1\beta_2}\epsilon_{\alpha_3\alpha_4}\epsilon_{\beta_3\beta_4}\epsilon \dots \{ \theta_{sol}^{\alpha_1},\theta_{sol}^{\beta_1} \} \{ \theta_{sol}^{\alpha_2},\theta_{sol}^{\beta_2} \} \{ \theta_{sol}^{\alpha_3},\theta_{sol}^{\beta_3} \} \dots \\
\label{kill3}
\xi' &=& 0 \\
\label{kill4}
\bar{\xi}'&=& 0.
\end{eqnarray}
$\bar{\xi}$ in (\ref{kill2}) contains all of the 4n $\theta^{\alpha A}$'s in Lorentz invariant way hence multiplying any $\theta^{\alpha A}$ yields zero. This solution is $\frac{1}{2}$ BPS, thus there is ${\cal N}=1$ SUSY while the trivial (diagonal) background is supposed to have ${\cal N}=2$ SUSY. It is also interesting to note that $n$ has to be finite in order for the above (\ref{kill2}) to be a Killing spinor. Should we take the limit $n \to \infty$ first, the Killing spinor (\ref{kill2}) is ill-defined hence half of the symmetries are broken. This case may rather corresponds to ${\cal N}=\frac12$ SUSY studied in \cite{Seiberg:2003yz}.


\section{10-dimensional case}

It is also straightforward to generalize the above solution to 10 dimensional case.
The action for the 10-dimensional IIB matrix model or so called IKKT model \cite{Ishibashi:1996xs} is defined by the following action:
\begin{eqnarray}
S &=& Tr(\frac{1}{4}[X^a,X^b][X_a,X_b] -\frac{1}{2}\bar{\theta}\Gamma^a[X_a,\theta]). \label{10Daction}
\end{eqnarray}
Here the indices $a$ and $b$ run from $0$ to $9$ and $\theta$ should be understood as Majorana-Weyl fermion in 10 dimension, though we work in Eucleadean space-time signature in the following.
The equations of motion derived from (\ref{10Daction}) are
\begin{eqnarray}
\label{10deoqX}
0 &=& [X^b,[X^a,X_b]] +\frac{1}{2}\{ \bar{\theta},\Gamma^a \theta \} \\
\label{10deoqTh}
0&=& [X^a, \Gamma_a \theta] .
\end{eqnarray}
Enhanced $N=2 $ supersymmetry transformations are given by
\begin{eqnarray}
\label{10dsusyX}
\delta X^a &=& i \bar{\xi}\Gamma^a \theta \\
\label{10dsusyTh}
\delta \theta^{\alpha}&=&\frac{i}{2}[X^a,X^b](\Gamma^{ab}\xi)^{\alpha}+\xi'^{\alpha}\vmi{ I}.
\end{eqnarray}
The commutator of two supersymmetries yields
\begin{eqnarray}
\label{10dalg}
[\delta_1,\delta_2]&=& \vmi{ I}(i\bar{\xi}_2\Gamma^a \xi'_1-i\bar{\xi}_1
                       \Gamma^a \xi'_2)
                       \partial_a +\delta_{U(N)gauge}(2\bar{\xi}_2 
                       \Gamma^a\xi_1 X_a) \nonumber \\
                   && + \mbox{eq. of motion for } \theta ,
\end{eqnarray}
in a shorthand.

Now, just as 4-dimensional case, let us adopt the following anstatz for $\theta$,
\begin{equation}
\theta_{sol} ^{a} = \sum_{A=1}^{2n} \theta^{aA} T^{A},
\end{equation}
where $\theta^{aA} $'s are 10-dimensional spinors.
Then the anti-commutator for $\theta$ becomes
\begin{eqnarray}
\{ \theta^{\alpha}, \theta^{\beta} \} &=& \sum_{AB} \theta^{\alpha A}\theta^{\beta B}
                                          f_{AB0}\vmi{ I} \nonumber \\
&\equiv& C^{\alpha \beta}\vmi{ I}.
\end{eqnarray}
In 10 dimension, the spinor structure of $C^{\alpha \beta}$ yields the following expansion,
\begin{eqnarray}
C^{\alpha \beta} &=& \frac{1}{16}(\Gamma^{a}
                     {\cal C})^{\alpha \beta}({\cal C}\Gamma_a)_{\gamma 
                     \delta } C^{\gamma \delta} \nonumber \\
                 &+& \frac{1}{32\cdot5!}(\Gamma^{a_1 \cdots a_5}
                     {\cal C})^{\alpha \beta}({\cal C}\Gamma_{a_5 \cdots 
                     a_1})_{\gamma \delta } C^{\gamma \delta},  \label{C10D}
\end{eqnarray}
where ${\cal C}$ is the charge conjugation matrix in 10 dimension. From (\ref{10deoqX}), (\ref{10deoqTh}), we can derive the following conditions as 4 dimensional case:
\begin{eqnarray}
\label{10dsolX}
X^{a A}&=& \bar{\eta}\Gamma^a \theta^A  \label{10DXsol}\\
\label{10dsolTh}
({\cal C}\Gamma^a)_{\alpha \beta}C^{\alpha \beta} &=& 0. \label{10DCcond}
\end{eqnarray}
Here we need to introduce one new Grassmann parameter $\eta$ since $\theta$ and $\bar \theta$ are not independent unlike four-dimensional case. Also note that eq. (\ref{10DCcond}) actually gives a constraint among $2n$  $\theta^{\alpha A}$'s.  

This solution has the killing spinor which is $16n$-th order with
respect to $C^{\alpha \beta}$. Thus we have ${\cal N}=1$ supersymmetry.
The commutation and anti-commutation relations of the solution are
\begin{eqnarray}
\{\theta^{\alpha}, \theta^{\beta} \} &=& C^{\alpha \beta}\vmi{ I} \\
\left[ X^a,\theta^{\alpha} \right] &=& (\bar{\eta} \Gamma^a)_{\beta} C^{\beta \alpha}\vmi{ I}\\
\left[ X^a, X^b \right] &=& \frac{1}{96}\bar{\eta}\Gamma^{a_1 a_2 a_3}\eta 
               ({\cal C}\Gamma_{\sp \sp a_3 a_2 a_1}^{ab})_{\alpha \beta}
               C^{\alpha \beta}\vmi{ I}. 
\end{eqnarray}

\section{Discussion}

We have presented a fermionic solution of the matrix model and shown that the solution yields the same commutation and anticommutation relations  as those of  noncommutative superspace. Let us comment on several aspects of the solution.

Firstly, it is worth mentioning that the present solution has a self dual structure. In fact, denoting the commutator of $X^{a}$ as a field strength,
\begin{equation}
F^{ab}  \equiv [X^{a}, X^{b}]
\end{equation}
and with the definition for the dual field strength
\begin{equation}
({\tilde F})^{ab} = \frac{i}{2}\epsilon^{ab}_{E} {}_{cd}F^{cd},  \qquad \epsilon^{0123}_{E}=i,
\end{equation}
the field strength for the solution $F_{sol}$ satisfy the (anti-) self dual condition,
\begin{equation}
F _{sol} = -{\tilde F}_{sol}.
\end{equation}
Since $F _{sol} $ contains $({\bar \theta})^{2}$, the would be instanton index for the solution, $Tr F {\tilde F}$ is zero. It would be interesting to clarify the relation between the present solution and the more general self dual solutions including the instanton solutions.

Secondly, the parameter $C^{\alpha \beta}$ for non(anti-)commutativity in our solution is a product of Grassmann variables. One may find this as an unsatisfying aspect of the solution. 
Therefore it would be interesting to observe that there is another solution of  (\ref{motionX}), (\ref{motionTh}) and (\ref{motionbTh}),
\begin{eqnarray}
\label{solTh2_1} 
\theta^1 &=& \Gamma^1 \otimes \vmi{ I}, \\
\label{solTh2_2}
\theta^2 &=& \Gamma^2 \otimes \vmi{ I}, \\
\label{solbTh2_1} 
\bar{\theta}^1 &=& (\Gamma^3 + i \Gamma^4) \otimes \vmi{ I}, \\
\label{solbTh2_2} 
\bar{\theta}^2 &=& (\Gamma^5 + i \Gamma^6) \otimes \vmi{ I}, \\
\label{solX2}
X^a &=& -i \theta \sigma^a \bar{\theta} \otimes \vmi{ I}, 
\end{eqnarray}
where $\Gamma^i$ are $SO(6)$ gamma matrix.
The above solution obeys Seiberg's noncommutativity relations 
(\ref{ferminoncom}) - (\ref{fermibar}). In particular,  we can set $C^{\alpha \beta}
=\delta^{\alpha \beta}$ in this case. And there is no killing vector like (\ref{kill2}) hence we have exactly ${\cal N}=\frac{1}{2}$ supersymmetry.
However, those matrices are not in the representation of $u(n)$, which means, unfortunately, we cannot derive the equations of motions (\ref{motionX}) - (\ref{motionbTh}) from a  (hermitian) matrix model action in the first place. 

$X^{a}_{sol}$ also contains  bi-Grassmann factor. This seems inevitable from the commutation relation (\ref{bosenoncom}) where the righthand side contains a product of Grassmann variables. 
It would be interesting to pursue the physical interpretation of this feature and the relation with the preceding work \cite{Schwarz:pf} , which we leave for future investigation.

\begin{center} \begin{large}
Acknowledgments
\end{large} \end{center}
We would like to thank Y. Kimura for his collaboration at the early stage of the present work.
We are also indebted to M. Hayakawa, N. Ishibashi, S. Iso, H. Kawai and F. Sugino for fruitful discussions. This work is supported in part by the Grants-in-Aid for Scientific Research (13135223) of the Ministry of Education, Culture, Sports, Science and Technology of Japan.

\end{document}